\documentclass{emulateapj}
\usepackage{graphicx}

\newcommand{\beq}	{\begin{equation}}
\newcommand{\eeq}	{\end{equation}}
\newcommand{\beqa}{\begin{eqnarray}}
\newcommand{\eeqa}{\end{eqnarray}}

\def\simlt{\lower.5ex\hbox{$\; \buildrel < \over \sim \;$}}
\def\simgt{\lower.5ex\hbox{$\;. \buildrel > \over \sim \;$}}

\font\tenbi=cmmib10 
\newfam\bifam  \textfont\bifam=\tenbi

\font\tenbr=cmbx10
\newfam\brfam  \textfont\brfam=\tenbr

\font\squinttenbi=cmbx10 at 9pt
\scriptfont\brfam=\squinttenbi

\def\vecnabla{
              \setbox1=\hbox{$\bigtriangledown$}
                           \raise.45ex\hbox{$\bigtriangledown$\hskip-.97\wd1
                           $\bigtriangledown$\hskip-.97\wd1
                           $\bigtriangledown$\hskip-.97\wd1}
                           \raise.47ex\hbox{$\bigtriangledown$}}

\def\rsun{\ifmmode {\rm R}_{\mathord\odot}\else $R_{\mathord\odot}$\fi}
\def\msun{\ifmmode {\rm M}_{\mathord\odot}\else $M_{\mathord\odot}$\fi}
\def\lsun{\ifmmode {\rm L}_{\mathord\odot}\else $L_{\mathord\odot}$\fi}

\def\tmb{\ifmmode {T_{\rm mb}^{13}(x,y,v)}\else $T_{\rm mb}^{13}(x,y,v)$\fi}

\begin{document}

\title{The Turbulent Origin of Outflow and Spin Misalignment in Multiple Star Systems}

\author{Stella S.~R.~Offner$^1$}
\author{Michael M.~Dunham$^2$}
\author{Katherine I. Lee$^2$}
\author{H\'ector G.~Arce$^3$}
\author{Drummond B.~Fielding$^4$}
\affil{$^1$Department of Astronomy, University of Massachusetts, Amherst, MA 01003 USA; soffner@astro.umass.edu \\
$^2$Harvard-Smithsonian Center for Astrophysics, Cambridge, MA 02138 USA \\
$^3$Department of Astronomy, Yale University, New Haven, CT, 06520, USA \\
$^4$Department of Astronomy, University of California, Berkeley, CA, 94720, USA}

\begin{abstract}
The protostellar outflows of wide-separation forming binaries frequently appear misaligned. We use magneto-hydrodynamic simulations to investigate the alignment of protostellar spin and molecular outflows for forming binary pairs. We show that the protostellar pairs, which form from turbulent fragmentation within a single parent core, have randomly oriented angular momentum. Although the pairs migrate to closer separations, their spins remain partially misaligned. We produce $^{12}$CO(2-1) synthetic observations of the simulations and characterize the outflow orientation in the emission maps. The CO-identified outflows exhibit a similar random distribution and are also statistically consistent with the observed distribution of molecular outflows. We conclude that observed misalignment provides a clear signature of binary formation via  turbulent fragmentation. The persistence of misaligned outflows and stellar spins following dynamical evolution may provide a signature of binary origins for more evolved multiple star systems.
\end{abstract}

\section{Introduction}

As many as half of all stars reside in binary or multiple star systems \citep{lada06,duchene13}.  Protostars and young stellar objects exhibit an even higher incidence of multiplicity  \citep{chen13,tobin16}. Thus, most stars appear to {\it form} with siblings. 

However, what this ubiquitous multiplicity implies about the initial conditions of forming stars remains debated. A variety of mechanisms have been proposed to explain multiple star formation \citep{tohline02}, but these theories are difficult to verify since imaging close binaries requires sub-arcsecond resolution and high-optical depth limits observations of the earliest star formation stages.  Numerical simulations suggest two main channels for multiplicity: turbulent fragmentation \citep{fisher04,goodwin04} and disk fragmentation \citep{adams89}. In the former scenario, turbulence in the natal core leads to multiple density enhancements, which independently collapse. The latter produces secondaries through gravitational instability within a massive accretion disk.

Binary separation provides one possible means of distinguishing between these mechanisms.  Turbulent fragmentation produces initial separations $>500$ AU \citep{Offner10}, while disk fragmentation gives separations $<$ 500 AU \citep{kratter10}. 
Indeed \citet{tobin16} find the separation distribution for Class 0 and Class I sources in Perseus is bimodal, exhibiting peaks at $\sim$100 AU and  $\sim$3000 AU. These scales are consistent with the predictions for disk fragmentation and turbulent fragmentation, respectively. However, dynamical evolution may quickly modify the separations, and \citet{Offner10} found that initially wide binaries migrated to close separations ($<200$ AU) in $\sim$0.1 Myr. If substantial orbital evolution occurs during the main accretion phase, which lasts $\sim 0.5$Myr \citep{dunhamppvi14}, the citet{tobin16} protostellar sample may not reflect the primordial separation. Indeed, the Class I protostars show little evidence for a peak at large separations.

Another possible means of distinguishing between the formation scenarios is outflow orientation. Binaries forming within the same accretion disk likely have common angular momenta and therefore aligned stellar spins, whereas binaries formed via turbulent fragmentation likely possess independent angular momentum vectors and, thus, have randomly oriented spins. It is not possible to directly measure the spin of an accreting protostar, however, the direction of the outflow, which is launched within a few stellar radii of the protostar \citep{pelletier92}, is believed to reflect the angular momentum of the protostar and inner accretion region. Since outflows span thousands of AU to a few parsecs, they provide a promising signpost for binary system origins.  A number of protobinary systems with misaligned outflows have been observed \citep{chen08,lee16}. A SMA survey of multiple protostellar systems by  \citet{lee16} found that outflow orientations of pairs with separations $>1000$ AU are statistically consistent with random or anti-aligned orientations. While misaligned binary outflows have previously been reported in numerical simulations \citep[e.g.,][]{Offner11}, they have not been explored in detail. 

The protostellar accretion disk orientation may also indicate the angular momentum direction. Recent observations have revealed a number of multiple systems with misaligned disks \citep{jensen14,salyk14,williams14}. These confirm that circumstellar gas in binary systems can have very different angular momenta. However, the observational statistics of both misaligned disks and outflows remain tentative, and numerical simulations have not explored how outflow or protostellar properties evolve in time for either binary formation scenario.

Here, we use radiation-magnetohydrodynamics (MHD) simulations to study the formation and evolution of binary systems formed via turbulent fragmentation. Our protostar formulation models protostellar outflow launching and allows us to follow both the protostellar spin and outflow orientation. A few prior MHD studies have explored outflow launching in tight-binary systems \citep{vaidya13,sheik15}, but this is the first MHD study, including feedback, of {\it binary formation.} 

\section{Numerical Simulations}\label{sim}

We perform the simulations using the ideal magnetohydrodynamics (MHD) adaptive mesh refinement (AMR) code {\sc orion}  \citep[e.g,][]{li12}. The simulations include self-gravity, magnetic fields, radiation in the flux-limited diffusion approximation, and protostellar feedback due to both protostellar luminosity and protostellar outflows \citep{Offner09, cunningham11}.  

The initial conditions and AMR parameters are identical to \citet{Offner14b} and differ only by the addition of a magnetic field. The simulations model an isolated core on a Cartesian grid and begin with a sphere of uniform density, 10 K gas and radius $R_c = 0.065$ pc, which is confined by a warm (1000 K), low-density ($\rho_c/100$) medium.  
The basegrid resolution is $64^3$, and the initial core is resolved by two AMR levels.  Additional levels are inserted when the density exceeds the Jeans condition for a Jeans number of $N_J=0.125$ \citep{truelove97}. We refine gas with a density gradient $\Delta \rho/\rho=$0.6 by at least two AMR levels to ensure the outflow-core interaction is well-resolved. We also refine on radiation energy gradients: $\Delta E_r/ E_r < 0.15$, so that the warm circumstellar region is well-resolved. When the Jeans condition is violated on the fifth level, a Lagrangian sink particle forms \citep{krumholz04}. This particle represents an individual forming star and follows a sub-grid model for protostellar evolution, including radiative feedback \citep{Offner09} and protostellar outflows \citep{cunningham11}.  However, we do not resolve protostellar disks, which would lie inside the $4\Delta x_{\rm min} \simeq 100$ AU sink particle accretion radius \citep{krumholz04}. Accretion occurs through a combination of infall, gravitational torques and numerical viscosity \citep[e.g.,][]{kratter10}. We adopt outflow boundary conditions, such that high-velocity unbound gas exits the domain.  

We initialize the gas velocities with a turbulent random field with power in wavenumbers $k=1-10$ and a divergence free (solenoidal) vector field. By construction our initial turbulent field has relatively low ratios of rotational to gravitational energy: $\beta < 0.01$. This initial turbulence damps until it is replenished by energy injected from the protostellar outflows. 

The initial magnetic field is uniform in the $z$ direction ($\vec B = B_0 \hat z$),  and its magnitude is similar to that of observed cores, which have mass-to-flux ratios of $\sim 2$ \citep{crutcher12}. 

At sufficiently high-resolution, protostellar outflows will self-consistently magnetically launch \citep[e.g.,][]{tomida15, sheik15}. However, achieving the observed velocities of $\sim 100$ kms$^{-1}$ requires $\sim R_\odot$ resolution, which is too computationally expensive to follow over long timescales.  Instead, we adopt a protostellar model 
specifying the collimation angle, $\theta=$0.01 radians, and wind launching fraction, $f_w = 0.21$ (21\% of the accreted material is ejected by the outflow). This efficiency, together with the launching velocity, produces momentum injection consistent with estimates from observed protostars \citep[see][]{cunningham11}. The collimation angle is set on the highest resolution cells, which are much smaller than the outflow extent. Thus, the effective outflow collimation is determined principally by interaction with the core envelope \citep{Offner11,Offner14b}.  Material is launched at a fixed fraction, $f_k=0.3$, of the Keplerian velocity: $v = f_k (GM_p/R_p)$, where $M_p$ and $R_p$ are the instantaneous protostellar mass and radius, respectively. 
The ``spin" of the protostar, which depends on the angular momentum of the accreted gas, determines the instantaneous direction of the outflow. \citet{fielding15} describe our angular momentum treatment in detail. 

We investigate turbulent fragmentation in cores ranging from $4\msun-8 \msun$ using a variety of turbulent seeds. The core masses are sufficiently large to experience fragmentation but not so large that the forming stars are high-mass, which would require consideration of ionization.  Thermal pressure is more dynamically significant in smaller cores ($\lesssim 2\msun$) with similar properties, and we find these rarely fragment. 

Table \ref{simprop} summarizes the simulation properties for the three fiducial core masses.  We perform twelve simulations in total, four for each core mass. One M4 and M8 simulation are evolved for $>0.1$ Myr after the formation of the primary, while the remainder run for 10~kyr. The shorter calculations allow us to probe the initial distribution of separations and orientations for a broader range of conditions. Altogether the simulations form 5 single stars, 5 binaries and 2 triples: 11 pairs in total.

\begin{deluxetable}{lccccc}
\tablecolumns{9}
\renewcommand{\tabcolsep}{0.07cm}
\tablecaption{Model Properties \label{simprop}}
\tablehead{ \colhead{Model\tablenotemark{a}} &  
  \colhead{$M_{\rm core}$($\msun$) } &
 \colhead{$B_z$($\mu$G)}&
  \colhead{$\mu_{\phi}$}&
  \colhead{$\sigma_i$(kms$^{-1}$) } & 
 \colhead{$\sigma_{f}$(kms$^{-1}$)  } }
\startdata
M4     & 4.0 & 41.2  &  2.5  & 0.52   & 0.60  \\ 
M6     &  6.0 & 61.8  & 2.5  & 0.62   & 0.56 \\
M8     & 8.0 &  82.4 &  2.5  & 0.73  & 0.62   
\enddata
\tablenotetext{a}{Model name, core mass, initial magnetic field, mass-to-flux ratio relative to the critical value, initial 3D velocity dispersion, and 3D velocity dispersion at the time of binary formation. } 
\end{deluxetable}

\section{Results}\label{results}

\subsection{Fragmentation}

The initial field orientation introduces asymmetry and causes collapse preferentially along the field lines, where magnetic support is absent. This produces a flattened turbulent structure in the $x-y$ plane. Consequently, these cores experience more fragmentation than the non-magnetized cores of \citet{Offner14b}, which predominately formed single stars.
The turbulence promotes the formation of small scale filaments, which are prone to Jeans-type filament fragmentation \citep{fischera12,pineda15}. 
Because the initial core has little net angular momentum, this filamentary sub-structure rather than strong rotation is responsible for the binary formation.

\begin{figure}[t]
\begin{center}
\includegraphics[scale=0.6, trim={0.4cm 0.2cm 0.1cm 0.1cm}, clip]{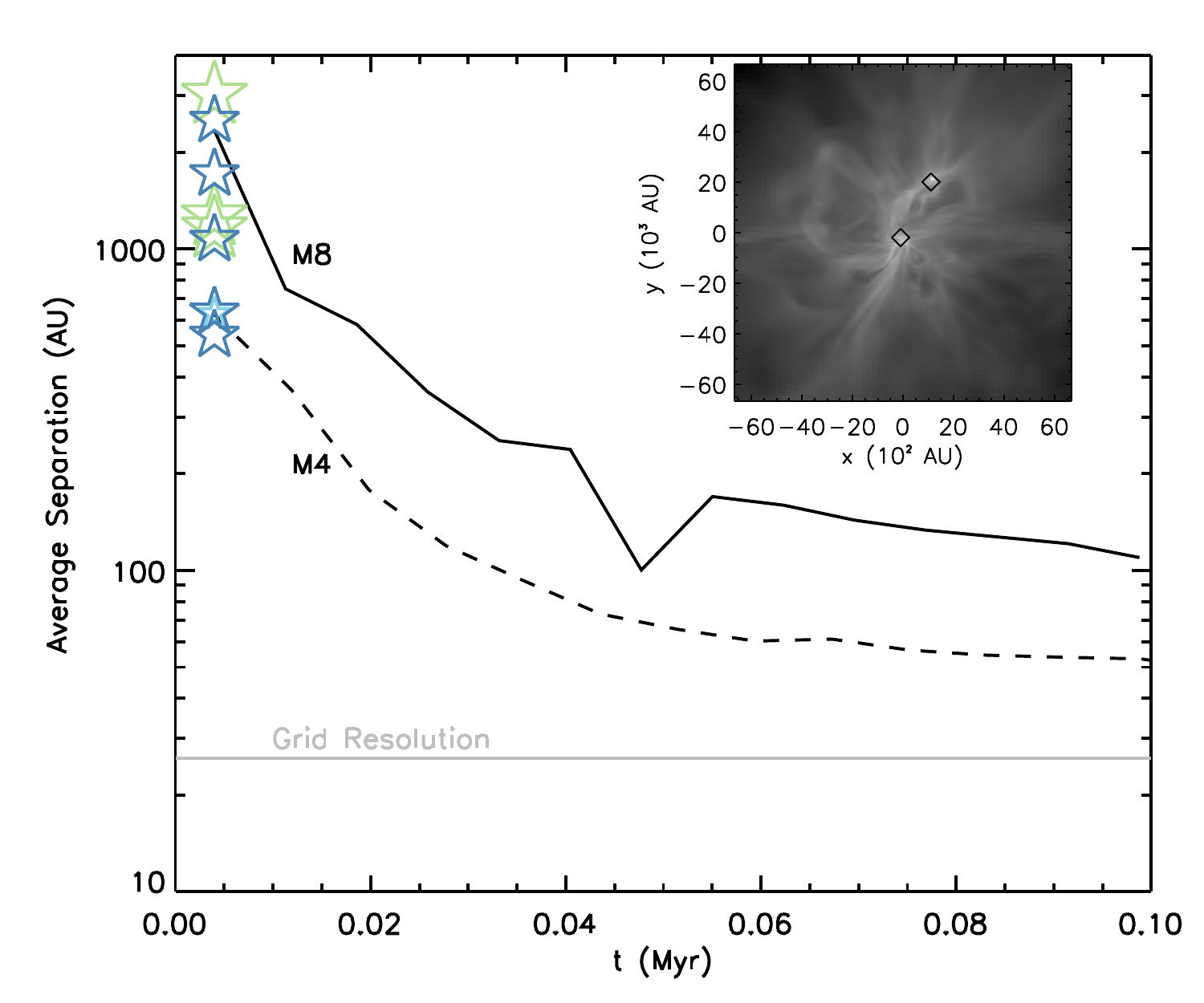}
\end{center}
\caption{Average separation versus time for the M4 and M8 binaries. The grey horizontal line indicates the minimum cell size. The stars show the initial pair separations for all cores  (4$\msun$ are small, light blue; 6$\msun$ are medium, dark blue; $8\msun$ are large, green).  The inset shows the log column density of M8 just after the formation of a secondary. Diamonds indicate the protostar positions. 
\label{massvst} }
\end{figure}

As Figure \ref{massvst} shows, 
the initial pair separations range from $\sim$600 to 3000 AU, which is consistent with observed wide separation core fragments and protobinaries  \citep{nakamura12,chen13,pineda15,lee16,tobin16}.   The inset illustrates that the fragmentation occurs within dense filaments created by the turbulence rather than within a massive accretion disk. The protostars in M4 begin with a separation of $\sim 600$ AU, which narrows to 50 AU over 0.05 Myr.  The pair in M8 evolve from 3000 AU to 100 AU over 0.1 Myr.  The rapid dynamical evolution and initial separations are similar to those in non-magnetized simulations of turbulent core fragmentation \citep{Offner10}. 

\subsection{Spin Alignment}

\begin{figure}[t]
\begin{center}
\includegraphics[scale=1]{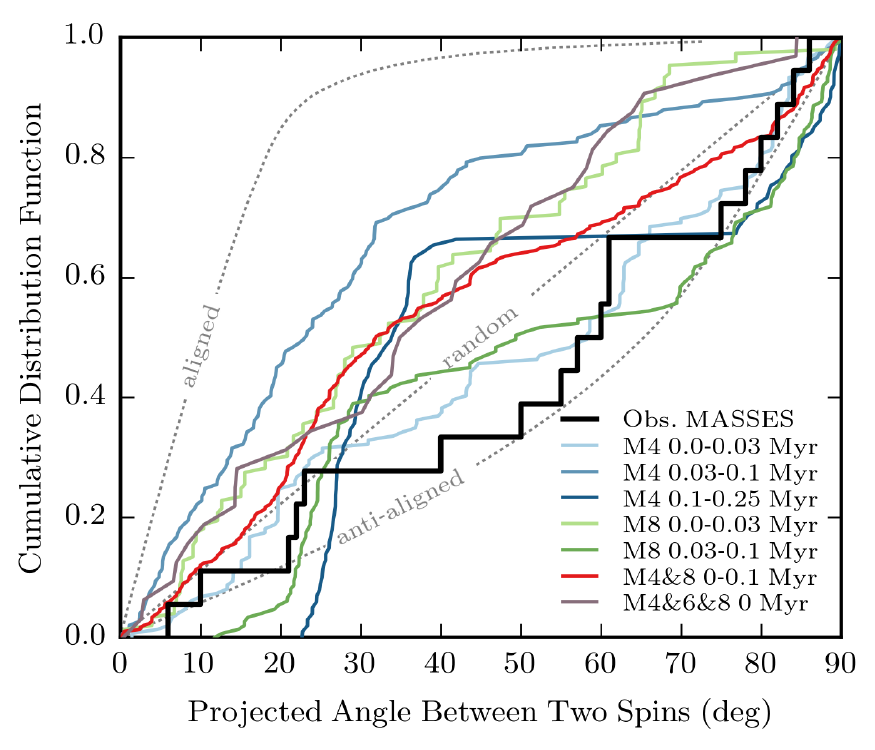}
\end{center}
\caption{CDF of the projected angle between the spins of protostellar pairs. Views along the $x$, $y$ and $z$ directions are treated as independent observations. The CDFs of projected orientation differences for a tightly aligned distribution (0-20$^{\circ}$, top dotted grey line), random distribution (middle dotted grey line), and preferentially anti-aligned distribution ($70-90^{\circ}$, bottom dotted grey line) were generated by 3D Monte Carlo simulations. The black histogram indicates the CDF of projected outflow angles for observed wide-binary pairs  \citep[MASSES,][]{lee16}.  
\label{simcdf} }
\end{figure}

 To investigate the protostellar spin alignment, we measure the projected angular difference of the spin viewed from the $x,~y$ and $z$ directions. We tabulate the angle differences every $\sim 400$ yrs following the formation of a secondary.  Figure \ref{simcdf} shows the cumulative distribution function (CDF) of angle differences for several age ranges.  The figure also displays the angles measured from outflow orientations in the Mass Assembly of Stellar Systems and their Evolution with the SMA survey  \citep[MASSES][]{lee16}. The protostellar systems in MASSES are predominately Class 0 objects, and so they are likely $0.15$ Myr or younger \citep{dunhamppvi14}.  Thus, we consider an interval just after formation ( $[0.0, 0.03]$ Myr) that represents early Class 0 sources, an interval spanning mid-Class 0 sources ($[0.03, 0.1]$ Myr), and an interval covering the Class 0/I transition  ([0.1, 0.3] Myr). We also combine the two datasets for the first 0.1 Myr, which we believe best corresponds to the evolutionary span of the MASSES sources. 
 
 The changing angular momentum of the accreting turbulent gas and the orbital interaction between the protostars causes significant spin evolution.
The angle differences are caused by changes in both spins rather than primarily one or the other. During the first interval ($\Delta = 0.03$ Myr), the simulated binary pairs are uncorrelated and are consistent with random orientations. This is consistent with their formation at wide separations from separate gravitational collapse events.  Over the next two time intervals the M4 spins become slightly more correlated, while the M8 spins become less correlated.

To quantify the CDF similarity, we perform a Kolmogorov-Smirnov (K-S) test. The K-S statistic gives 1 minus the confidence level at which the null hypothesis that the samples were drawn from the same parent distribution can be ruled out. We find K-S statistics between the MASSES observations and simulations of 10$^{-3}$, 0.50, and 0.004 for M4, M8 and the combined sample, respectively, over the first 0.1 Myr, and K-S statistics of 0.91, 0.21, and 0.75 for the first 0.03 Myr.   The K-S statistic between the MASSES observations and initial spin projections is 0.07.  Not all times and masses are statistically consistent; distributions exhibiting slightly less alignment agree better with the observations. 

\begin{figure*}[t]
\begin{center}
\includegraphics[scale=0.8, trim={0cm 0.9cm 0cm 1cm}, clip]{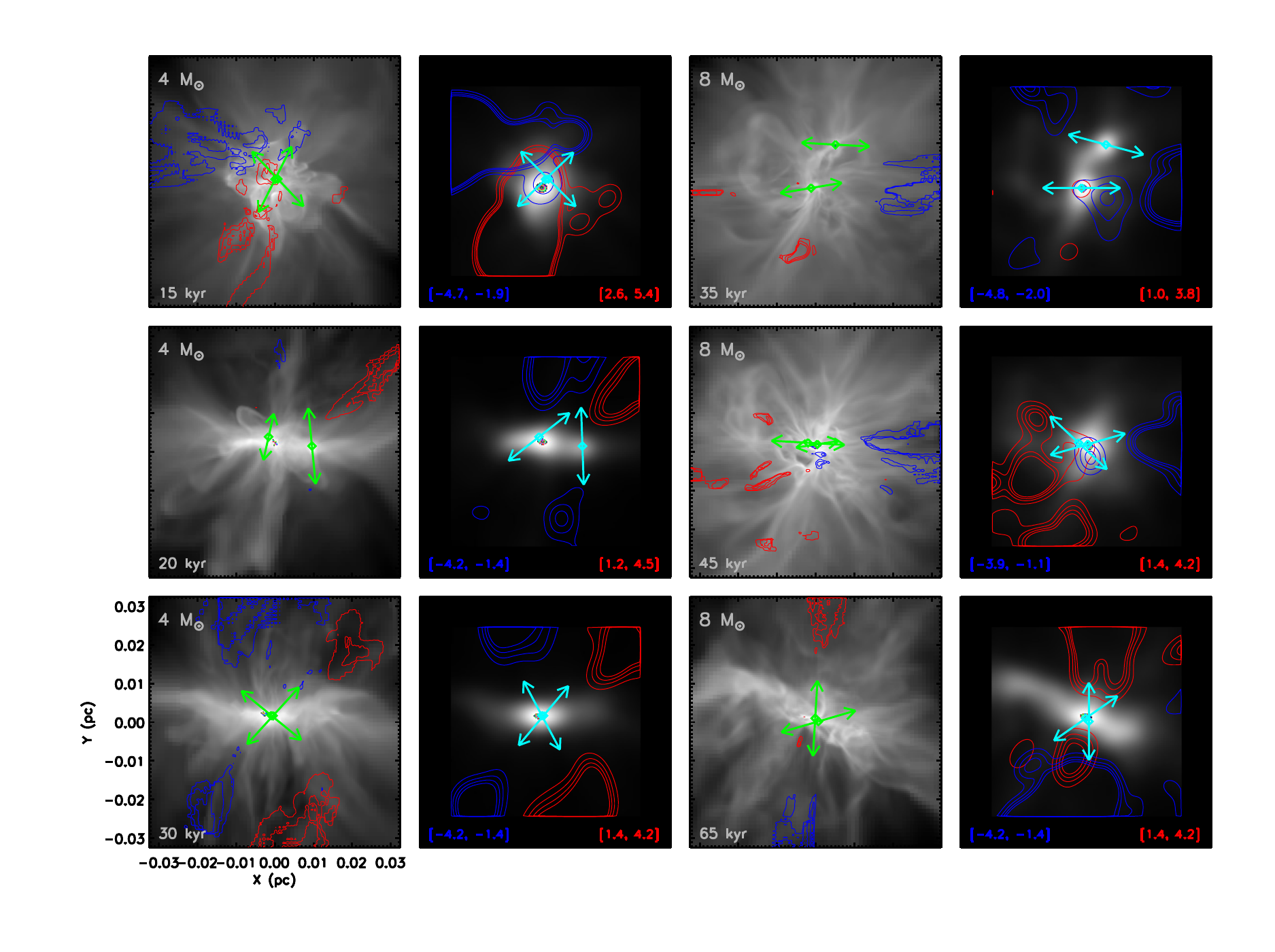}
\end{center}
\caption{Column density and projected red and blue shifted velocities for various M4 (left column) and M8 ( right column) snapshots.  Green arrows (left panel of each pair) indicate the projected protostellar spin directions; cyan arrows (right) display the visually identified outflow directions. The left panel of each pair shows contours of the integrated synthetic $^{12}$CO(2-1) emission overlaid on log column density. The contours are 60\%, 72\% and 86\% of the peak emission. The age of the primary appears in the lower left. The right panels show both the column density and emission convolved with a 4'' beam assuming the source is at 250 pc. The contours are $2\sigma, 4\sigma, 6\sigma$ and $8\sigma$ assuming 0.2 K noise per $0.5 {\rm km s^{-1}}$ channel, which is typical of the MASSES data. The integrated velocity range appears at the bottom. The range is adjusted slightly to include most of the outflow emission and exclude the core gas and empty channels. 
\label{cooutflow} }
\end{figure*}

\subsection{CO Outflow Alignment}

While protostellar spins reflect the angular momentum and serve as a proxy for outflow direction, they are not observable. In order to compare with the MASSES data, we use {\sc radmc-3d}\footnote{\url{http://www.ita.uni-heidelberg.de/~dullemond/software/radmc-3d/}}, a line radiative transfer code, to produce synthetic maps of $^{12}$CO (2-1). We perform the radiative transfer using the non-LTE Large Velocity Gradient approximation. The molecular excitation and collisional data are taken from the Leiden atomic and molecular database \citep{schoier05}.  

We first flatten the AMR data to 256$^3$ resolution over a region of 0.065 pc ($\Delta x=$52 AU). To convert the simulation mass densities to CO densities, we adopt an abundance of $10^{-4}$ CO per H$_2$ \citep[e.g.,][]{Offner14b}. The CO abundance for gas with temperatures $>800$ K is set to zero to reflect CO dissociation in low-density gas and the ionized jet. The CO abundance is also assumed to be zero for densities $n_{\rm H_2}>2 \times 10^4 ~{\rm  cm}^{-3}$ to account for CO freeze-out onto dust grains. Each synthetic cube spans $\pm 10$ km/s and has a channel width of 0.08 km~s$^{-1}$. We produce emission cubes for three orthogonal views for outputs separated by 0.01 Myr and protobinary ages $\leq 0.1$ Myr.  

Figure \ref{cooutflow} displays a subset of the CO outflows at different viewing angles and times.   The synthetic outflows show a range of morphologies. Some appear well-collimated  (middle left), while others exhibit poor collimation (top left). The outflow lobes are usually asymmetric, which is consistent with observations \citep{lee15}. For most outputs, the outflows are distinct and do not combine to form a single collective outflow as in \citet{peters14}. 

We identify outflows from integrated maps of the blue and red-shifted emission and, following \citet{lee16}, measure the projected outflow angle difference manually. Uncertainties in the angle measurement are $\pm 5 \deg$.  We exclude maps that do not exhibit two distinct outflows (e.g., Figure \ref{cooutflow} bottom right). Due to the close protostar proximity and projection effects, only 18 and 20 of the 30 views for M4 and M8, respectively, would likely be detectible. This tends to remove pairs with more aligned outflows, since these are harder to distinguish. However, this exclusion is consistent with MASSES, which by design only includes resolved ($>1000$ AU), distinguishable outflows.  

\begin{figure}[t]
\begin{center}
\includegraphics[scale=0.95]{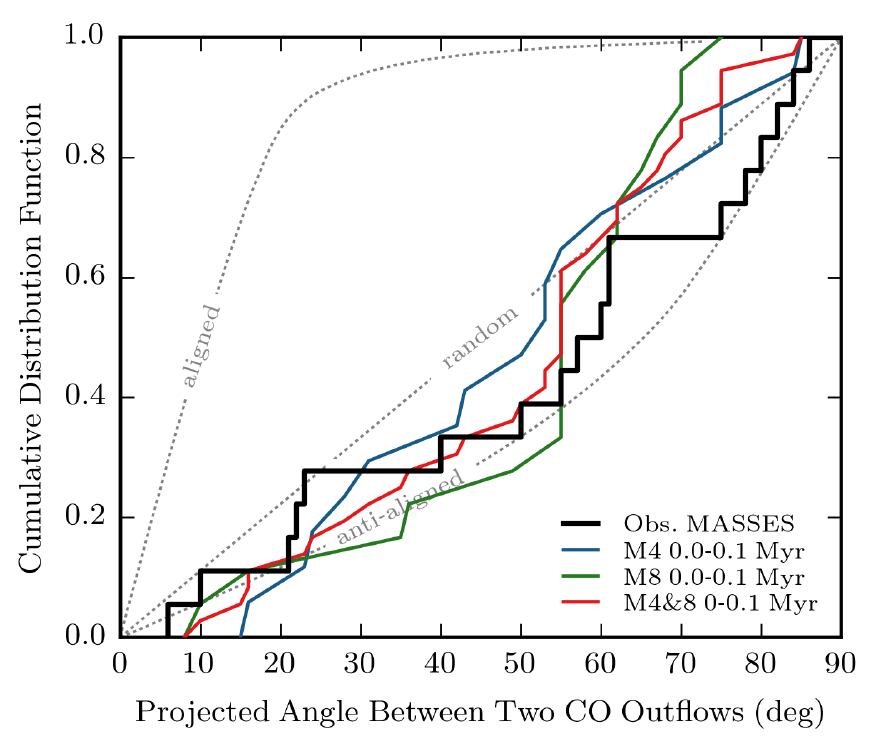}
\end{center}
\caption{CDF of the projected angle between outflow pairs identified from synthetic CO(2-1) maps. Grey lines display the MASSES data. Simulation outflow orientations are identified for protobinary ages $\leq 0.1$ Myr for views along the $x$, $y$ and $z$ axes, where the outputs are spaced in intervals of 0.01~Myr.
\label{cocdf} }
\end{figure}

Figure \ref{cocdf} shows the CDF of the projected orientations of CO-identified outflows for binary ages $\le$0.1 Myr. The CDF is statistically consistent with both the random and misaligned distributions of outflow orientations. A K-S test returns a statistic of $0.57$ for the combined synthetic CO outflow CDF and the observations, which indicates strong statistical consistency. 

The CDF of synthetic CO angle differences have a K-S statistic of $0.08$ when compared to the protostellar spin distribution for the same time range. This indicates that the projected spins are statistically consistent with the orientations inferred from the synthetic CO outflow maps; however, as illustrated by Figure \ref{cooutflow}, the projected spins are often slightly offset from the molecular outflow.


\section{Discussion} \label{conclude}

\subsection{Statistical Considerations}

The statistical agreement between the synthetic spin and outflow distributions and the observations is promising but tentative. Both simulations and observations require larger sample sizes  for robust conclusions.  MASSES is the first survey of wide-separation protostellar systems with arcsecond resolution, uniform sensitivity and completeness. However, MASSES contains only 19 pairs. Additional sources are needed to statistically discriminate between a random and anti-aligned CDF. Further observations of outflow and disk alignment versus separation could help constrain the degree of misalignment for close separation binaries.

Likewise, simulations with more complete sampling of physical conditions, including different magnetic and turbulent properties, should be performed in future work. Our simulations suggest outflows can change direction rapidly. Many observed outflows appear constant on parsec scales
; however, other sources are candidates for significant directional change \citep{hsieh16}. Tight-binary interactions may also produce visible jet deflection \citep{fendt98,vaidya13}. Additional comparisons of observed and simulated outflows on larger scales is necessary to explore the impact of angle variation on outflow morphology. Despite these limitations, the increasing pace of discoveries of misaligned disks and outflows in wide multiple systems underscores that a common physical mechanism is at work. 

\subsection{Magnetic Fields and Binary Formation}

The ideal MHD approximation assumes the gas and field are well-coupled.  Strong and well-coupled fields remove angular momentum through magnetic braking and, thus, reduce accretion disk sizes or even eliminate them altogether \citep{lippvi14}.  The details of the disk sizes do not impact our results, however, since binary formation occurs though turbulent core fragmentation rather than disk fragmentation \citep[see also][]{li10}.

Simulations with ideal MHD do exhibit more efficient angular momentum transport than those with nonideal treatments. Consequently, it is possible that including effects such as ohmic dissipation and ambipolar diffusion will produce {\it larger} differences between protostellar spins and even less aligned outflows than we find here.  However, turbulence, which promotes small scale reconnection of magnetic field lines, also counteracts magnetic breaking and even permits disk formation in ideal MHD simulations \citep{lippvi14}.


A few 3D simulations have explored the impact of nonideal effects on accretion and outflows. For example, \citet{sheik15} find that magnetic diffusivity impacts the accretion rate and may enhance outflow velocities. \citet{tomida15} find differences due to dissipation but only at densities exceeding $10^{-11}$ g cm$^{-3}$, which is beyond our resolution. Global star formation simulations including nonideal MHD effects are necessary to determine the true impact on binary formation, accretion and outflow properties.

\subsection{Spin-Outflow and Spin-Orbit Misalignment}

Our outflow model assumes that the instantaneous outflow launching direction is identical to the angular momentum vector of the protostar. If the inner disk ($<10R_\odot$), which is not resolved here, has a different orientation than the protostellar spin, this assumption breaks down. The degree of correspondence between the accretion disk and protostellar rotation during the protostellar phase, when the spin cannot be measured, is unknown. However, the inner disk where the protostellar outflow launches is expected to have a similar rotation to the protostar since magnetic and gravitational torques 
star spin and disk over time \citep{lai11,batygin13,lai14}. Other observations demonstrate that the star spin and debris disk inclinations are aligned \citep[$|i_*-i_d|<10^\circ$,][]{watson11,greaves14}, which supports coincident outflows and stellar spins.

In contrast, exoplanet systems frequently display misalignment between the stellar rotation and orbital plane of hot Jupiters \citep{winn10}: ``spin-orbit" misalignment.  
If these planetary orbits reflect the final disk orientation, this implies that the disk and star angular momenta were at one point misaligned.  

A variety of mechanisms have been proposed to explain spin-orbit misalignment, including dynamical interactions between stars and/or planets \citep{nagasawa08}, the chaotic star formation environment during the accretion phase \citep{bate10}, and perturbations from a binary companion \citep{thies11, batygin12,lai14}. 
\citet{fielding15} used hydrodynamical simulations of protostars forming in a turbulent clump to explain the spin-obliquity of hot Jupiters, which they propose results from turbulent motions during the accretion phase. Accretion of turbulent gas with differing net angular momenta is also the mechanism that creates the misaligned protobinary systems here.  Turbulent origins naturally explains the lack of correlation between protobinary spins and the misalignment between observed outflow pairs. 

\subsection{Implications for Multiple Star Formation}

The persistence of the simulated binary spin misalignment suggests that the coincidence, or lack thereof, of more evolved binary pair spins may reveal their formation channel. We predict that systems with significantly misaligned stellar spins, independent of separation, may originate from turbulent fragmentation.  Thus, spin misalignment could be a relic of formation. Indeed, misaligned spins in solar-type binary systems appear common \citep{hale94}, with wider separation binaries having larger inclination offsets. When a disk is present, misalignments between the disk and spin may also occur as a result of an eccentric binary companion \citep{batygin12, lai14}. In principle, this mechanism may also cause stellar spins to be misaligned. However, the relevant time-scale is 
much longer than the disk orbital time and protostellar phase \citep{lai11}.    

Triple star systems, which are generally hierarchical (one close pair with a more distant tertiary), have necessarily undergone dynamical evolution erasing their primordial configuration. Measuring the spins of the individual members could reveal whether such systems formed via disk fragmentation, turbulent fragmentation or a combination of the two.  \citet{hale94} found that equatorial inclinations of hierarchical triples are frequently misaligned and the misalignment distribution does not decrease with increasing separation. This may occur because dynamical interactions cause misalignments 
or because the inclinations are lingering signposts of turbulent formation at wide separations.

\acknowledgements 
We acknowledge support from NASA grant NNX15AT05G (SSRO), NASA grant NNX13AE54G (MMD), NSF GRFP under Grant No. DGE 1106400 (DF), a Submillimeter Array postdoctoral fellowship (MMD), the Yale University High Performance Computing Center and the Massachusetts Green High Performance Computing Center.


\end{document}